# Concentration-Dependent Restructuring of Ionic Liquid Micelles Induced by an Anionic Surfactant

Devansh Kaushik[a*], Sajal K. Ghosh[a], and Syed M. Kamil[a*]

[a] *Department of Physics, School of Natural Sciences, Shiv Nadar Institution of Eminence, NH 91, Tehsil Dadri, Uttar Pradesh -201214, India*

## Abstract

The self-assembling behaviour of ionic liquids in aqueous solution is important for understanding their physicochemical properties and for their industrial applications. While the influence of ionic liquids on surfactant micellization has been widely studied, much less attention has been given to how surfactants affect the aggregation of ionic liquids, particularly when the surfactant concentration is below its critical micelle concentration (CMC). In this work, we examine the effect of the anionic surfactant sodium dodecyl sulfate (SDS), introduced at sub-CMC concentration, on the micellization of 1-methyl-3-octylimidazolium chloride ($[OMIM]^+[Cl]^-$) in aqueous solution maintained above the IL CMC, using surface tension measurements, theoretical analysis, and coarse-grained molecular dynamics simulations. We find that at low SDS concentrations ~2 mM, SDS inserts smoothly into the pre-existing IL micelles, producing stable mixed micelles with favourable IL–SDS interactions. When the SDS concentration approaches ~4 mM, the micelles exhibit distinct changes in their internal dynamics, reflected in deviations in the thermodynamic parameters. Beyond this point, as more SDS is added, the system reorganizes and forms stable mixed micelles again, now containing a higher fraction of SDS but still enriched in IL. The synergistic behaviour is quantified using Clint and Rubingh's models, and simulations supports the structural transitions, showing variations in micelle size, aggregation number, and radial distribution functions. This work demonstrates that SDS acts as an effective modulator of IL aggregation, providing mechanistic insight into IL–surfactant co-assembly.

*Corresponding Authors: Dr. Syed Mohammad Kamik (kamil.syed@snu.edu.in), Devansh Kaushik (dk912@snu.edu.in)

## 1. Introduction

Amphiphilic molecules such as surfactant and ionic liquids contain both hydrophobic and hydrophilic moieties in the molecular structure. Amphiphiles reduce the surface tension of water as they orient at the air-water interface by disrupting the hydrogen bonding between the water molecules at the interface [1,2]. In an aqueous solution, above the critical micelle concentration (CMC), amphiphiles aggregate in the form of micelles, or higher hierarchical structures [3]. Amphiphiles are widely used in products such as detergents, cosmetics, paints, pharmaceuticals, personal care and in many other industrial processes, including oil recovery and emulsification [4–6]. The formation of these structures depends on the molecular shape and size, the nature of headgroup interactions, and the hydrophobicity of the surfactant tail [7,8]. Furthermore, the self-assembly of surfactants are sensitive to system-specific factors and can be regulated through adjustments in physical parameters such as temperature and pH. Due to their widespread application and interesting physiochemical behaviour, both industrial and academic research communities have their interest in the field.

The amphiphilic ionic liquids (ILs) are the salts that have melting points below 100 °C and consist entirely of organic cations with hydrophobic tails and complementary inorganic anions [9]. ILs allow precise selection of cations and anions which ensures the control over their physio-chemical properties, hence known as designer solvents. Due to their unique properties of high thermal stability, low flammability, non-volatility, excellent biodegradability, high solvation ability, and tunability, they are known as environmentally friendly alternatives of surfactants [10]. Hence, ILs are being explored in the various fields such as catalysis, batteries, biotechnology, biomass dissolution, wound healing, drug delivery, and cosmetic industries. Like most amphiphilic molecules, ILs form self-assembled structures when placed in aqueous solution [11].

Numerous studies have demonstrated that the addition of co-surfactants, organic solvents, and electrolytes can modulate inter-surfactant interactions, thereby enabling control over mesophase formation and the critical micelle concentration (CMC) [12–14]. Consequently, surfactant mixtures are often preferred over individual surfactants because of their synergistic effects [15,16]. The synergism arises from the interactions between differently charged surfactant head groups (cationic/ anionic), which alters the adsorbance to interface, surface tension, as well as the formation of mixed micelles [17,18]. Micellization is primarily an entropy-driven process governed by the hydrophobic effect. In aqueous solution, water molecules form highly ordered “iceberg” structures around the hydrophobic tails of surfactant monomers to preserve hydrogen bonding within the bulk water. Upon micelle formation, these structured water molecules are released into the bulk, resulting in a positive entropy change ($\Delta S>0$). However, in mixed micellar systems (e.g., nonionic/ionic, anionic/cationic, or zwitterionic/ionic mixtures), micellization deviates from this ideal behaviour [19]. Such deviations arise from attractive or repulsive interactions between dissimilar headgroups, variations in hydrophobic interactions, counterion effects, and other molecular factors.

Kotsi et al. [20] reported synergistic interactions between a nonionic surfactant (tristyrylphenol ethoxylates) and an anionic surfactant (sodium benzenesulfonate, C10–C13) during micellization and adsorption at the air–water interface. They observed that premixed surfactant solutions required up to 15 hours to reach equilibrium, whereas systems in which the surfactants were added sequentially equilibrated within approximately 40 minutes. They also used theoretical modelling to predict the attractive interactions in the mixed micelles and at the interface layers, while the supramolecular assemblies in the bulk and interface are preferentially enriched in non-ionic surfactants. These findings underscore how co-adsorption and micellization of structurally distinct surfactants – especially when their individual CMCs differ by multiple orders of magnitude – allow formulation strategies to tune interfacial packing, adsorption dynamics and micellar stability in complex amphiphile systems. Farahani *et al.* demonstrated that cationic ionic liquids can cooperatively interact with amphiphilic drug molecules, forming stable mixed micelles with markedly reduced critical micelle concentrations [21]. They used different theoretical models (Clint, Rubingh, Motomura and Rodenas) to calculate the interaction parameters, which suggested the formation of stable mixed micelles in synergistic state [22,23]. Their work highlights how electrostatic and hydrophobic complementarities govern mixed IL–drug micellization, providing a broader framework for understanding IL–surfactant co-assembly. For this study, they have used antidepressant drugs, imipramine hydrochloride (IMP), amitriptyline hydrochloride (AMT) and ionic liquid as 1-dodecyl-3-methylimidazolium bromide). Their results of mixed antidepressants with ionic liquids indicated the decrease in CMC value with addition of ionic liquid in the mixture. On the similar front, a work by Mitra et al. have showed the effect of ionic liquids on the mesophases of a surfactant sodium dodecyl sulfate (SDS) [24]. It was shown that the addition of varied concentrations of short chained ILs can induce the transition from hexagonal to lamellar and back to hexagonal, which was quantified using the small angle x-ray scattering (SAXS) results.

Numerous studies have examined the influence of ionic liquids on the aggregation behaviour and critical micelle concentration (CMC) of surfactants [24–26]. In contrast, the reverse, how surfactants restructure IL micelles, remains largely understudied. Keeping this in mind, the present study explores the effect of anionic surfactant (SDS) on the micellization behaviour and CMC of ionic liquid (1-methyl-3-octyl imidazolium chloride ($[OMIM]^+[Cl]^-$). The study employs CMC calculation by using a surface tensiometer. The experimental data were analysed using the Clint and Rubingh models of mixed micellization to calculate the relevant parameters. These modes are based on the Regular Solution Theory (RST), and have been used widely towards understanding the synergism in mixed micellar systems. Additionally, the formation of mixed micelles was confirmed using the coarse-grained molecular dynamics using Martini force field, where the radius of gyration of the micellar aggregates and the diffusion of SDS molecule in them were studied.

## 2. Materials and Methods

### 2.1. Materials

The surfactant, sodium dodecyl sulfate (SDS, 99% purity) and the ionic liquid (IL), 1-methyl-3-octyl imidazolium chloride ($[OMIM]^+[Cl]^-$, 98% purity) were purchased from Sigma Aldrich (USA). Figure 1(a) and (b) shows the chemical structure of SDS and $[OMIM]^+[Cl]^-$, respectively. De-ionized water (Milipore, 18MΩcm) was used to prepare the IL-SDS aqueous solutions. Throughout the experiments, the concentration of IL was kept constant (350 mM, above CMC), while the SDS concentration was varied systematically between 0 and 6 mM (below CMC of SDS). To attain the equilibrium in mixed system, the aqueous mixture was left undisturbed overnight at room temperature, which were then used for further experiments.

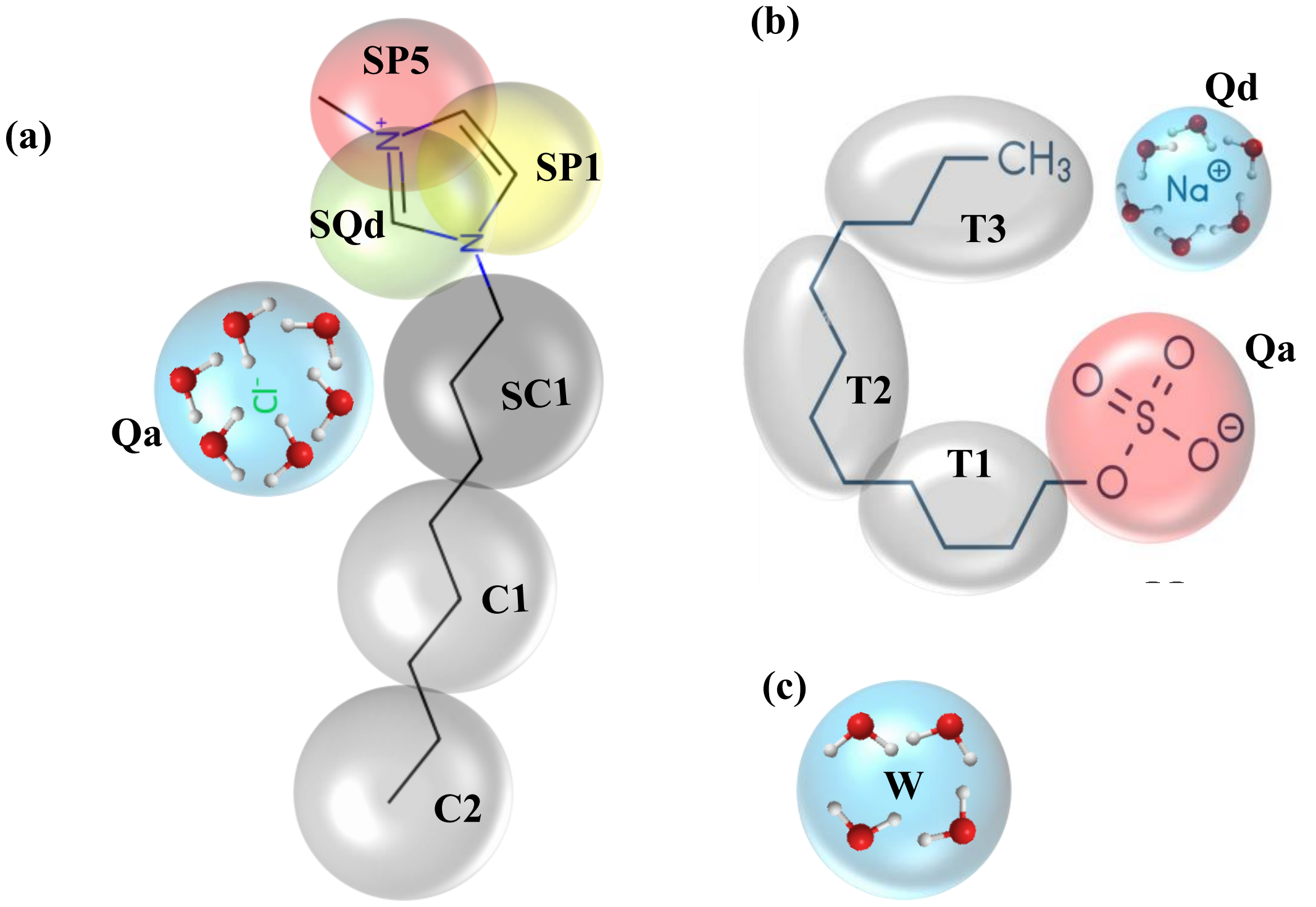

***Figure 1.*** *The chemical structure of (a) sodium dodecyl sulfate (SDS), and (b) 1-methyl-3-octyl imidazolium chloride (*$[\text{OMIM}]^+[\text{Cl}]^-$*) and (c) water, with their respective CG models overlayed on each molecular structure.*

## 2.2. Methods

### 2.2.1. Critical Micellar Concentration

A force balance Sigma 701 (Attension, Biolin Scientific), equipped with a platinum Wilhelmy plate, was used for the surface tension measurements. The value of surface tension (σ) can be calculated based on the equation,

$$\boldsymbol{\sigma} = \frac{\boldsymbol{F(N)}}{\mathbf{2}(\boldsymbol{t_p} + \boldsymbol{w_p})\ \mathbf{cos}\ \boldsymbol{\theta}} \tag{1}$$

where, $F(N)$ is the force acting on the plate, which is measured, $t_p$ is the thickness of the plate, $w_p$ is the width of the plate, and $\theta$ (°) is the contact angle between the plate and the liquid. As surface tension measurement is very sensitive to impurities, the platinum Wilhelmey plate was rinsed with de-ionized water followed by acetone before exposing to flames. Further, the surface tension of de-ionized water (~72 mN/m) was measured at room temperature [25]. This process was repeated before every independent measurement, to ensure the accuracy. To calculate the critical micelle concentration (CMC), 30 ml of aqueous solution of IL was placed in the vessel and the surface tension at the initial concentration was measured. Further, the solution was diluted sequentially and the surface tension was measured after 5 minutes of each dilution. The curve between surface tension and concentration was used to measure the CMC value of the mixture. All the surface tension measurements were carried out at room temperature (25 °C) and ambient pressure.

### 2.2.2. MD simulations

To probe the structure and molecular assembly, coarse grained molecular dynamics (CG-MD) integrated with Martini force field was employed [26]. The simulations were done using LAMMPS [27]. The ionic liquid concentration was fixed at 350 mM, while the concentration of Sodium dodecyl sulfate was varied between 0 and 6 mM (0, 2, 4, and 6 mM). The water molecule is modelled using P4 bead with each bead containing 4 water molecules. 10% of BP4 beads were added to avoid the freezing tendency of martini water at regular temperatures [26]. Sodium and chloride ions are represented with a charged CG bead with a hydrogen donor (Qd bead) and acceptor (Qa bead), respectively, that contain six implicit water molecules. The ionic liquid was modelled as 3:1 mapping, while SDS was modelled as 4:1 mapping. The force field parameters for IL [28] and SDS [29] were taken from literature. Figure 1(a)-(c) shows overlayed coarse-grained models used through the simulations. The simulation was initially run for 10 ns for equilibrating the system, further 10 ns were used as production run at NPT ensemble. The time step for the simulation was set at 10 fs and the temperature was kept at 300 K. The supplementary details are presented in section 1 of Supplementary Information (SI).

The viscosity of liquids can be determined from equilibrium molecular dynamics simulations using the Green–Kubo formalism, which connects macroscopic transport coefficients to microscopic fluctuations. Specifically, the shear viscosity is obtained from the time integral of the autocorrelation function of the off-diagonal elements of the pressure tensor [30],

$$\eta = \frac{V}{k_B T} \int \langle P_{\alpha\beta}(0) P_{\alpha\beta}(t) \rangle dt$$

where V is the system volume, T is the temperature, $k_B$ is the Boltzmann constant, and $P_{\alpha\beta}$ ($\alpha\beta$=xy, xz, yz) are the shear stress components. The three independent off-diagonal terms are averaged to improve statistical reliability. This approach provides a direct link between molecular-level stress fluctuations and the macroscopic viscous response of the fluid, and is widely applied to study complex liquids such as ionic liquids and surfactant assemblies.

## 4. Results

### 4.1. Critical Micelle Concentration – Ideal Mixing

Micelles form in aqueous solutions of surfactants or ionic liquids once the concentration exceeds a threshold value known as the critical micelle concentration (CMC). Multiple studies have quantified the $CMC$ using UV-Vis, ionic conductivity and surface tension measurements [31,32]. Here, the surface tension of pure ionic liquid ($[OMIM]^+[Cl]^-$) and in presence of surfactant (SDS) was measured to obtain the $CMC$ of pure and mixed aqueous systems (see Figure S1).

Figure 2(a) shows the $CMC$ variation of IL-SDS system as a function of surfactant (SDS) concentration, where the decrease in $CMC$ is observed. Researchers have employed various theoretical models, including those of Clint, Rubingh, Motomura, and Rodenas, to determine mixed micelle composition and calculate their thermodynamic parameters based on the individual CMCs of the constituent surfactants [22,23,33,34]. The parameters give insight towards the ideal and non-deal nature of mixing. These thermodynamic parameters can also be utilized to predict the prevailing interactions in the system.

According to Das et al. [35], the critical micelle concentration of a mixed surfactant system ($CMC_{Ideal}$) is related to the CMCs of its individual components ($CMC_i$) through the following equation,

$$\frac{\mathbf{1}}{\boldsymbol{CMC_{Ideal}}} = \sum\left[\frac{\boldsymbol{\alpha_i}}{\boldsymbol{f_i \times CMC_i}}\right] \tag{2}$$

In the above equation, $\alpha_i \left(= \frac{IL}{IL+SDS}\right)$ and $f_i$ denote the mole fraction and activity coefficient, respectively, of $[OMIM]^+[Cl]^-$ (i =1) and SDS (i = 2) in the mixed micelle. In the ideal case, where the mixed micelles behave ($f_i$=1) and there are no interactions between the components, the CMC of the mixture can be calculated using the Clint equation given by,

$$\frac{\mathbf{1}}{\boldsymbol{CMC_{Ideal}}} = \frac{\boldsymbol{\alpha_1}}{\boldsymbol{CMC_1}} + \frac{\mathbf{1} - \boldsymbol{\alpha_1}}{\boldsymbol{CMC_2}} \tag{3}$$

Here, $CMC_{Ideal}$ is the ideal CMC of the mixture, $CMC_1$ and $CMC_2$ are the CMCs of the pure IL and SDS, respectively, and $\alpha_1$ and $1-\alpha_1$ are the bulk mole fractions of IL and SDS. Any difference between $CMC_{Ideal}$ and the experimental CMC of the mixture ($CMC_{Mix}$) are caused by the interactions between the two molecular species. A positive deviation ($CMC_{Mix} > CMC_{Ideal}$) indicates repulsive interactions, whereas a negative deviation ($CMC_{Mix} < CMC_{Ideal}$) reflects attractive interactions. As shown in Figure 2(b), the CMC of the IL/SDS mixture exhibits a negative deviation from the ideal value, indicating non-ideal mixing with an attractive interaction between the components. These interactions likely result from hydrophobic effects combined with electrostatic attraction between the oppositely charged headgroups.

### 4.2. Micellar Mole Fraction – Non-Idealistic Mixing

In general, mixtures of ionic or non-ionic surfactants with similar structures can provide reasonable theoretical predictions of their behaviour in micellar phase through ideal solution theory which has been discussed in the previous section. However, mixture of surfactants with different structures and significantly different $CMC$ usually exhibits more complex and non-ideal behaviour that cannot be predicted by ideal solution theory. Therefore, the behaviour of mixed micelles is approximated using the Rubingh's model for regular solution theory [36], using the equation,

$$\left(X_1^{Rub}\right)^2 \ln\left[\frac{\alpha_1\, CMC_{Mix}}{X_1^{Rub}\, CMC_1}\right] = \left(1 - X_1^{Rub}\right)^2 \ln\left[\frac{(1-\alpha_1)\, CMC_{Mix}}{\left(1 - X_1^{Rub}\right) CMC_2}\right] \quad (4)$$

where $X_1^{Rub}$ is the micellar mole fraction of IL in a mixed micelle, while all other symbols have their usual meaning. The micellar mole fractions of IL ($X_1^{Ideal}$) for the ideal system were obtained using equation 3 based on Motomura approximation [37,38], using the equation,

$$X_1^{Ideal} = \frac{\alpha_1 CMC_2}{\alpha_1 CMC_2 + \alpha_2 CMC_1} \quad (5)$$

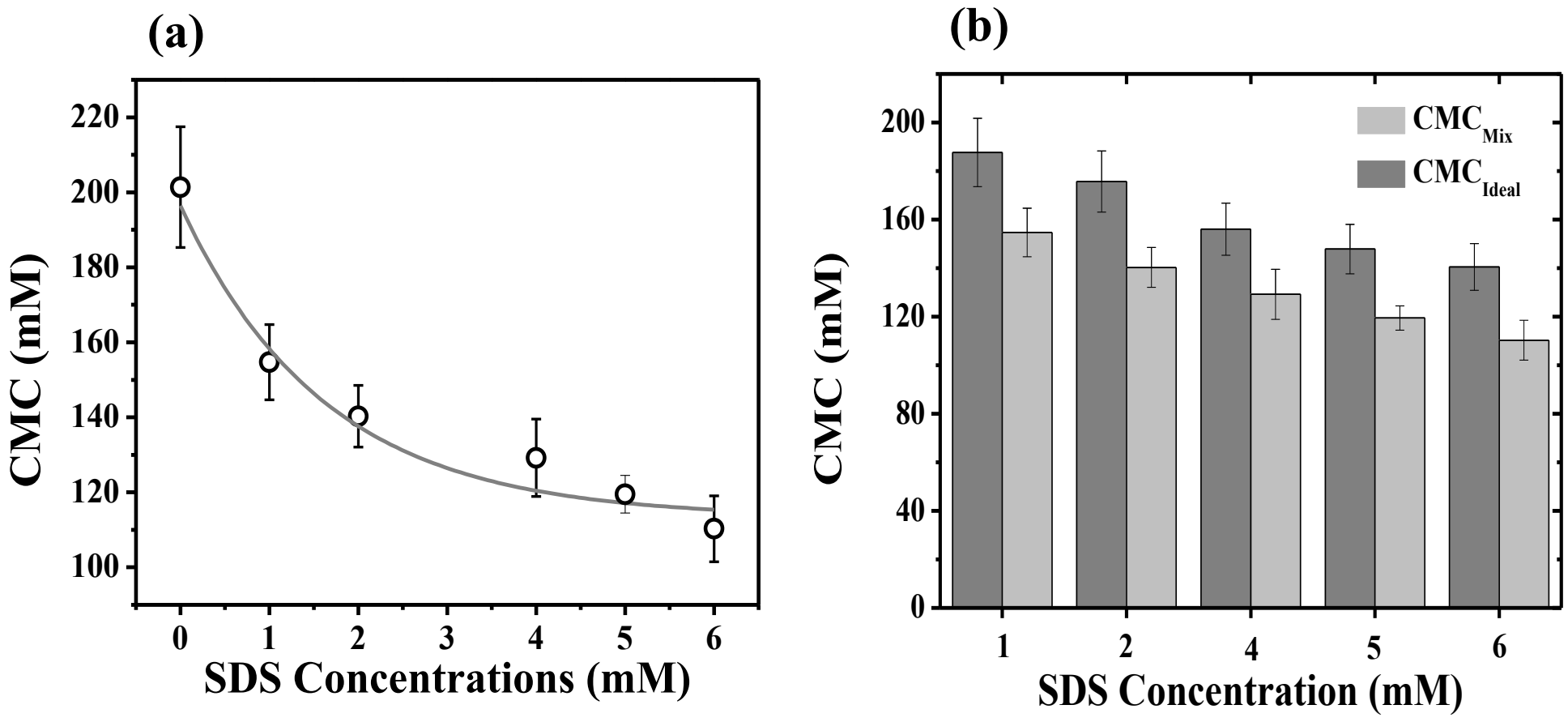

***Figure 2. (a)*** *Change in the critical micellar concentration of ionic liquid, 1-methyl-3-octyl imidazolium chloride ($[OMIM]^+[Cl]^-$) as a function of added surfactant, sodium dodecyl sulfate (SDS).* ***(b)*** *The comparison between the experimental critical micellar concentration ($CMC_{Mix}$) and the CMC calculated based on the Clint's model ($CMC_{Ideal}$).*

The deviation of $X_1^{Rub}$ from $X_1^{Ideal}$ for a given mole fraction would indicate the non-ideality of the mixed micelles. As the $X_1^{Ideal}$ is higher compared to $X_1^{Rub}$, it suggests that the molecules of specie 1 (IL) are being replaced by specie 2 (SDS). Figure 3(a) shows that the mole fraction of IL in the 'mixed micelle' decreases with the increasing SDS concentration. This suggests the formation of IL rich micelles, where SDS molecules penetrates the pre-existing IL micelles. Here, the partitioning of SDS molecule is facilitated by the hydrophobic interactions between the tail group of both amphiphiles, where the charge neutralization can further lead to the formation of stable micellar aggregates.

The $X_1^{Rub}$ values calculated from equation 5, has been used to calculate the interaction parameter ($\beta$), using the equation [36],

$$\beta = \ln\left(\frac{\alpha_1\, CMC_{Mix}}{X_1^{Rub}\, CMC_1}\right) \cdot \frac{1}{\left(1 - X_1^{Rub}\right)^2} \quad (6)$$

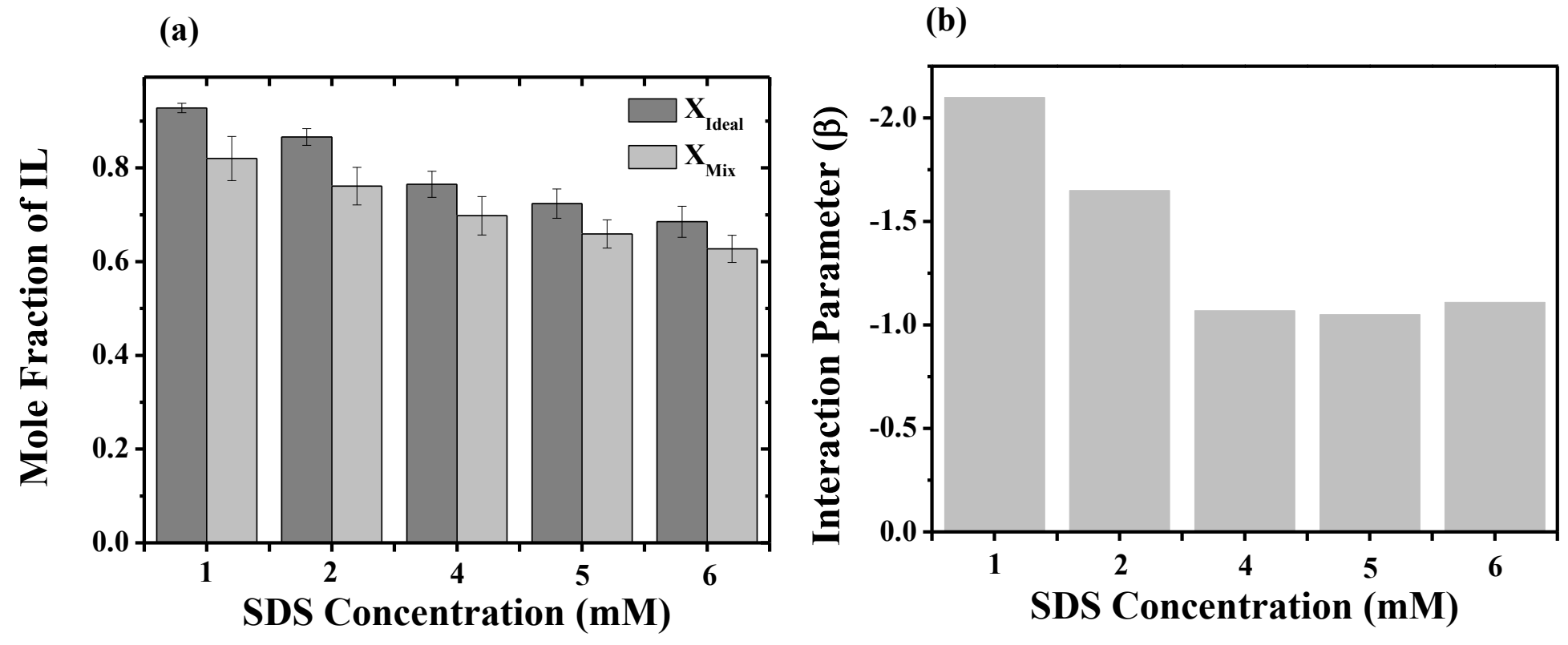

***Figure 3.*** *(a) Variation of the micellar mole fraction of [OMIM]$^+$[Cl]$^-$ in mixed micelles as a function of SDS concentration, calculated using Rubingh's model. The decreasing trend indicates progressive incorporation of SDS molecules into the micelles. (b) Interaction parameter (β) versus SDS concentration, showing negative values across all compositions. The least negative β at 4 mM SDS suggests a transient destabilization of the micellar structure, while more negative values at lower and higher SDS concentrations indicate stronger synergistic interactions and enhanced micelle stability.*

The $\beta$ values represent an information about the type and strength of interaction between the two amphiphiles. A zero value shows that the interaction between components is approximately similar, while negative and positive means synergistic and antagonistic effects between the components caused by attractive or repulsive interaction, respectively. Figure 3(b) shows the values of interaction parameter ($\beta$), as a function of increasing SDS concentration. The obtained values of $\beta$ are negative over the entire range of SDS concentration, which shows the favourable mixed micelle formation. The type of interaction between the amphiphiles can include electrostatic, van der Waals, steric and hydrogen bonding between two groups in different molecules [39]. Here, the electrostatic attraction and hydrophobicity are the major driving forces for the stable mixed micelle formation. Interestingly, the $\beta$ value at 4 mM SDS concentration corresponds to the least stable micelle, post which the stability again increases at 5 and 6 mM. This anomalous behaviour is intriguing and need to be explored.

The activity coefficient informs about the non-ideality and strength of intermolecular interaction between two components in a mixed micellar system. According to the regular solution theory, activity coefficients ($f_1^{Rub}$ for IL and $f_2^{Rub}$ for SDS) can be obtained using the values of $\beta$ and $X_1^{Rub}$ calculated using equation 4 and 6 [36],

$$\boldsymbol{f_1^{Rub} = \exp\left[\beta\left(1 - X_1^{Rub}\right)^2\right]} \qquad (7)$$

$$\boldsymbol{f_2^{Rub} = \exp\left[\beta\left(X_1^{Rub}\right)^2\right]}$$

The value of activity coefficient, $f = 1$, shows the ideal behaviour of resembling surfactants. Figure 4 shows the value of activity coefficient is less than unity for both mixed components at all concentrations of SDS, confirming the interaction between IL and SDS. Additionally, the figure 4 shows the higher values of activity coefficient of IL ($f_1^{Rub}$) compared to the activity coefficient of SDS ($f_2^{Rub}$), confirming the partition of SDS molecule into the IL rich micelles. While Clint's and Rubingh's provide a macroscopic understanding of micellar stability and interactions, molecular simulations enable visualization of the mixed micelles and their stability.

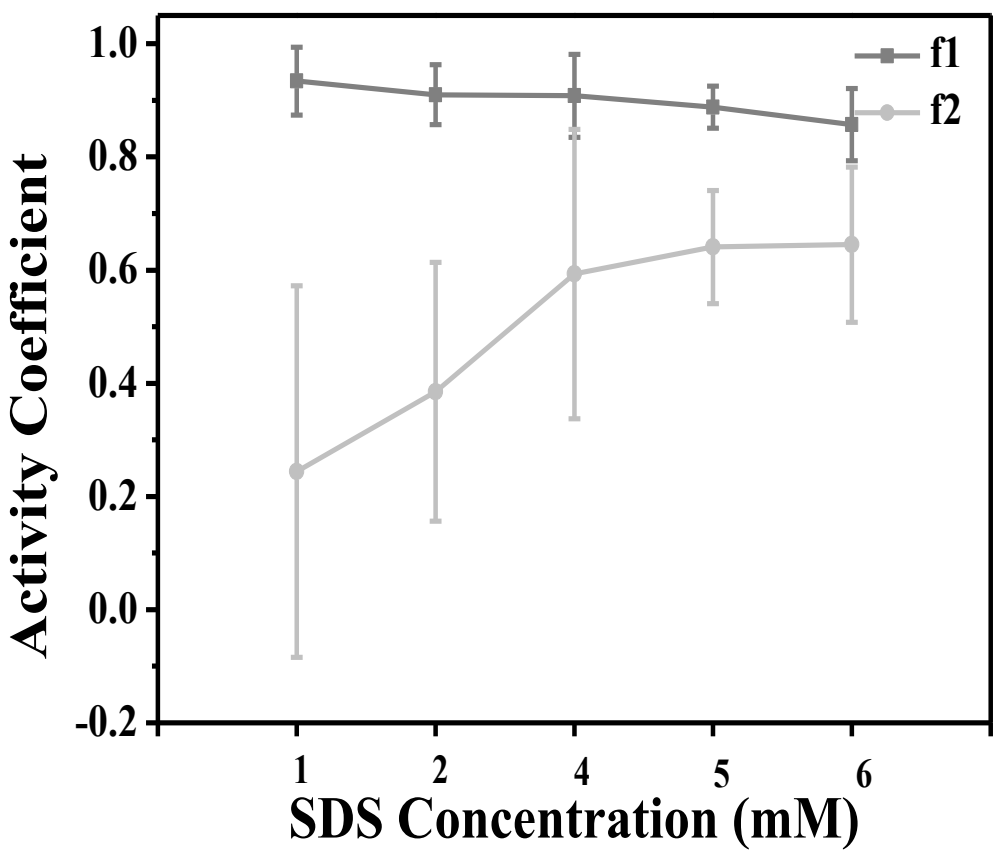

***Figure 4.*** *The plot of activity coefficient ($f_1^{Rub}$ (IL) and $f_2^{Rub}$(SDS)) as a function of SDS concentration, in the mixed micelles of IL-SDS.*

**4.3. Molecular Dynamics**

To visualize the structural changes in the mixed mesoscopic phase, the coarse-grained molecular dynamics (CG-MD) technique with Martini force field has been employed. The concentrations were similar to those studied experimentally. The CG model of the IL, water and SDS molecules are shown in Figure 1. Snapshots of the equilibrated systems containing 350 mM 1-octyl-3-methylimidazolium chloride ($[OMIM]^+[Cl]^-$) in the absence and presence of SDS are shown in Figure 5(a–d). In these systems, the concentration of the ionic liquid was kept constant at 350 mM, while the SDS concentration was varied from 0 to 6 mM, corresponding to 0, 2, 4, and 6 mM SDS in panels (a–d), respectively. Here, the black bead depicts the carbon tail of IL molecules (SC1, C1 and C2 beads, see Figure 1(b)), whereas yellow beads are the tails of SDS molecules (T1, T2 and T3 beads, see Figure 1(a)), while the red beads represent the head of SDS molecule (H), the head of ILs and water beads have been omitted for a better representation. The equilibrated mixed IL-SDS system confirmed the formation of mixed micelles, where the SDS molecules partition into the IL-rich aggregates, confirming the experimental prediction of formation of mixed micelles. The inset of Figure 5(b) shows a zoomed image of SDS partitioned into the IL aggregate.

As the SDS concentration increases, the structural properties of the IL micelles change in a non-monotonic manner. The average radius of gyration ($R_g$) varies from 13.28 ± 3.64 Å at 0 mM SDS to 16.16 ± 3.5, 12.85 ± 1.23, and 14.54 ± 1.54 Å at 2, 4, and 6 mM SDS, respectively. Similarly, the number of beads in the largest three micelles, fluctuates from 284 ± 20 beads at 0 mM SDS to 392 ± 14, 234 ± 19 and 294 ± 6, beads at 2, 4, and 6 mM SDS, reflecting complex redistribution of surfactant molecules and micellar restructuring with increasing SDS concentration. Incorporation of SDS reduces the charge on the micellar surface, promoting the formation of more tightly packed aggregates and resulting in a decrease in both the critical micelle concentration (CMC) and the aggregation number ($N_{agg}$). The non-monotonic behaviour of the IL-rich micelles is also evident in the radius of gyration of the aggregates, with a clear threshold observed at 4 mM SDS. For analysis, only aggregates with an average radius of gyration equal to or larger than the atomistic size of the surfactant molecules was considered. Density profiles of the largest aggregates formed are shown in Figure 5(e)–(h), where the red curve shows the presence of SDS in the IL rich micelles. At low SDS concentration (2 mM), SDS is localized within a narrow region, but its distribution broadens as the concentration increases. The molecular simulations reveal a non-monotonic evolution of the micellar radius of gyration and aggregation number with increasing SDS concentration, characterised by initial growth, followed by a collapse near 4 mM SDS, and partial re-growth at higher SDS levels. This behaviour reflects a redistribution of surfactant molecules between coexisting micellar populations rather than a simple monotonic size change. In experiments, such microscopic restructuring is manifested indirectly through thermodynamic observables, including the mixed CMC, micellar composition, and interaction parameter β, which exhibit their strongest deviations near the same SDS concentration.

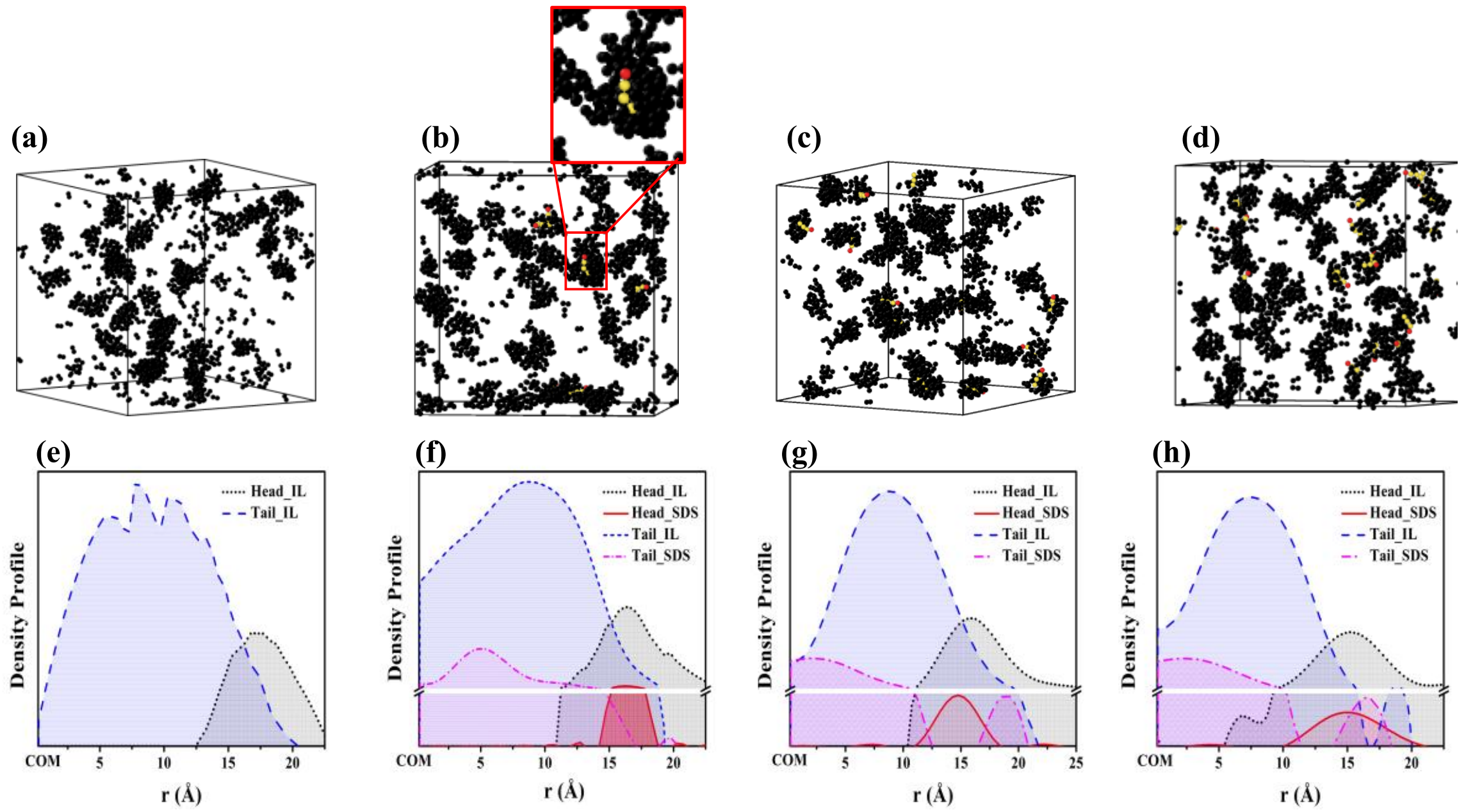

***Figure 5.*** *The aggregates formation of IL-SDS system and the respective density profile of largest micellar aggregate in presence of* ***(a, e)*** *0 mM,* ***(b, f)*** *2 mM,* ***(c, g)*** *4 mM and,* ***(d, h)*** *6 mM SDS using coarse grained molecular dynamics.*

Micelles are dynamic in nature, and the motion of the hydrophobic tail of SDS molecule within the aggregates can provide useful information regarding the micellar stability. To examine this behaviour, the trajectory of SDS molecule was tracked for 200 ps in the equilibrated system, and the molecular localization have been shown in Figure 6. The green spheres in the figure 6 represents the terminal tail bead (C2) of the SDS molecule, indicating the instantaneous position of the hydrophobic tail recorder at the end of 200 ps. Figure 6(a) shows that at 2 mM SDS concentration, the SDS molecule fluctuates inside a stable aggregate. As the concentration of SDS increases to 4 mM, Figure 6(b) shows the heightened motion of SDS molecule, which can be related to the lowest interaction parameter ($\beta$), which forms the least stable micelles. With the further increase in the surfactant concentration, the motion of SDS molecule is restricted again inside the micellar aggregate (see Figure 6(c)), suggesting the formation of stable micelle. This increase in stability can also be seen through the increase in interaction parameter at 6 mM SDS concentration. The re-stabilization observed at 6 mM SDS likely corresponds to a critical threshold of SDS incorporation, where electrostatic, hydrophobic, and packing forces reach a balance, enabling the formation of uniform, energetically favourable mixed micelles.

To quantify these observations, the diffusion length $l_d$ was calculated from the trajectory path of the SDS tail bead (see Supporting Information for details). During the analysis, outliers corresponding to bead displacements greater than 75 Å were excluded (distance corresponding to the average diameter of two mixed micelles). The average diffusion lengths were 27.87 Å, 37.36 Å, and 31.84 Å at 2, 4, and 6 mM SDS, respectively. The larger $l_d$ at 4 mM reflects increased SDS mobility and reduced micellar stability, whereas the smaller values at 2 mM and 6 mM are consistent with more confined motion within relatively stable aggregates.

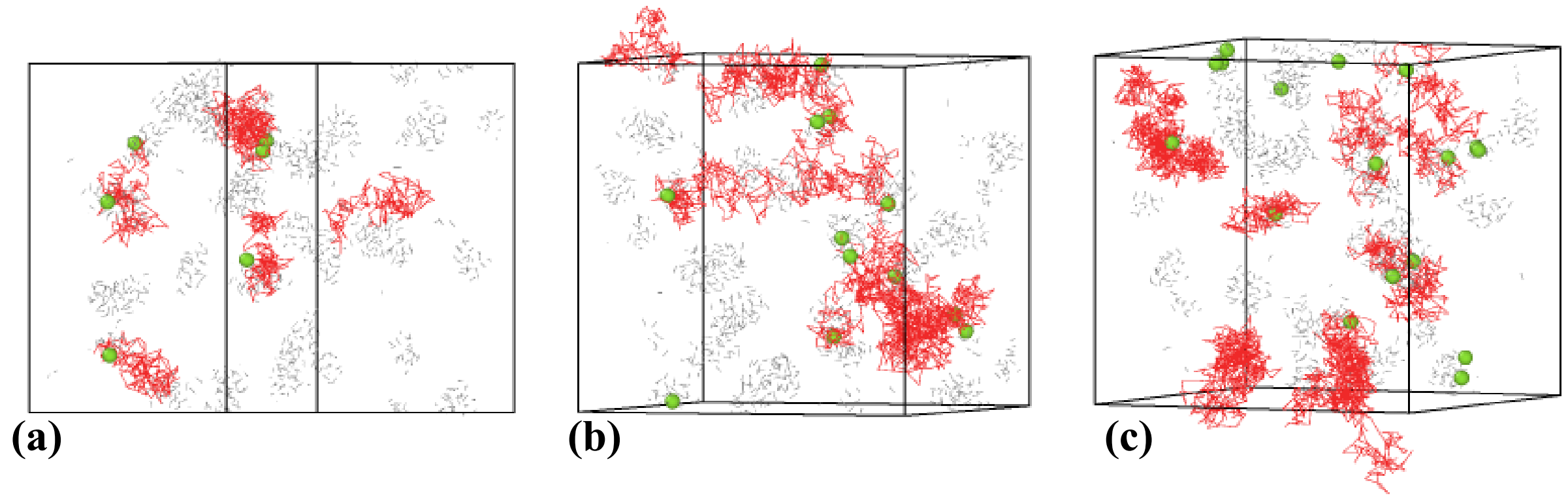

***Figure 6.*** *The figure shows the trajectory of SDS molecule in the aqueous solution of IL and SDS (a) 2 mM, (b) 4 mM, and (c) 6 mM of the aqueous solution of IL and SDS for 200 pico-seconds. The red lines represent the trajectory of the surfactant molecule during the simulation window. For clarity, the terminal tail bead (T3) of the SDS molecule is highlighted as green spheres, indicating the instantaneous position of the hydrophobic tail recorder at the end of 200 ps. The trajectories illustrate the local motion of SDS molecules within ionic-liquid–rich clusters.*

The radial distribution function (RDF) combined with viscosity offers structural and dynamical insights towards the effect of SDS on micellization of IL (see Figure 7). Figure 7(a) represents the RDF between the C2–C2 beads (See Figure 1(a)), corresponding to IL (zoomed image of first peak is depicted in the inset), across varying SDS concentrations. At 0 mM SDS, the RDF peak is sharp and intense, indicating strong tail–tail correlations and well-packed IL-rich micelles. As SDS is gradually incorporated (2 to 6 mM), the intensity of the primary RDF peak decreases with progressive increase in SDS concentration. However, the same peak position throughout implies that while the degree of ordering diminishes, the characteristic intermolecular spacing among IL tails remains intact. Thus, SDS incorporation weakens tail–tail correlations but does not completely alter the micellar core architecture. This is consistent with the experimental observation of decreasing IL mole fraction in mixed micelles, which reflects the gradual replacement of IL molecules by SDS.

The impact of SDS on micelle dynamics and stability is further captured by the zero-shear viscosity shown in Figure 7(b). The viscosity increases upon addition of 2 mM SDS, indicating enhanced micelle stability and tighter packing due to favourable IL–SDS interactions, likely driven by electrostatic attraction and hydrophobic compatibility. However, at 4 mM SDS, viscosity drops sharply, aligning with the observed minimum in the interaction parameter (β) (Fig. 3(b)) and the highest degree of SDS mobility within the micelles as stated previously. This suggests a critical SDS concentration around 4 mM, where the micellar structure becomes transiently destabilized, likely due to excessive surface crowding, electrostatic imbalance, or suboptimal tail–headgroup packing. Interestingly, the viscosity rises again at 6 mM SDS, indicating reorganization into more stable mixed micelles with better structural integration of SDS. Together, the RDF and viscosity results underscore a concentration-dependent modulation of micellar organization, where SDS first reinforces, then destabilizes, and finally restabilizes the aggregates.

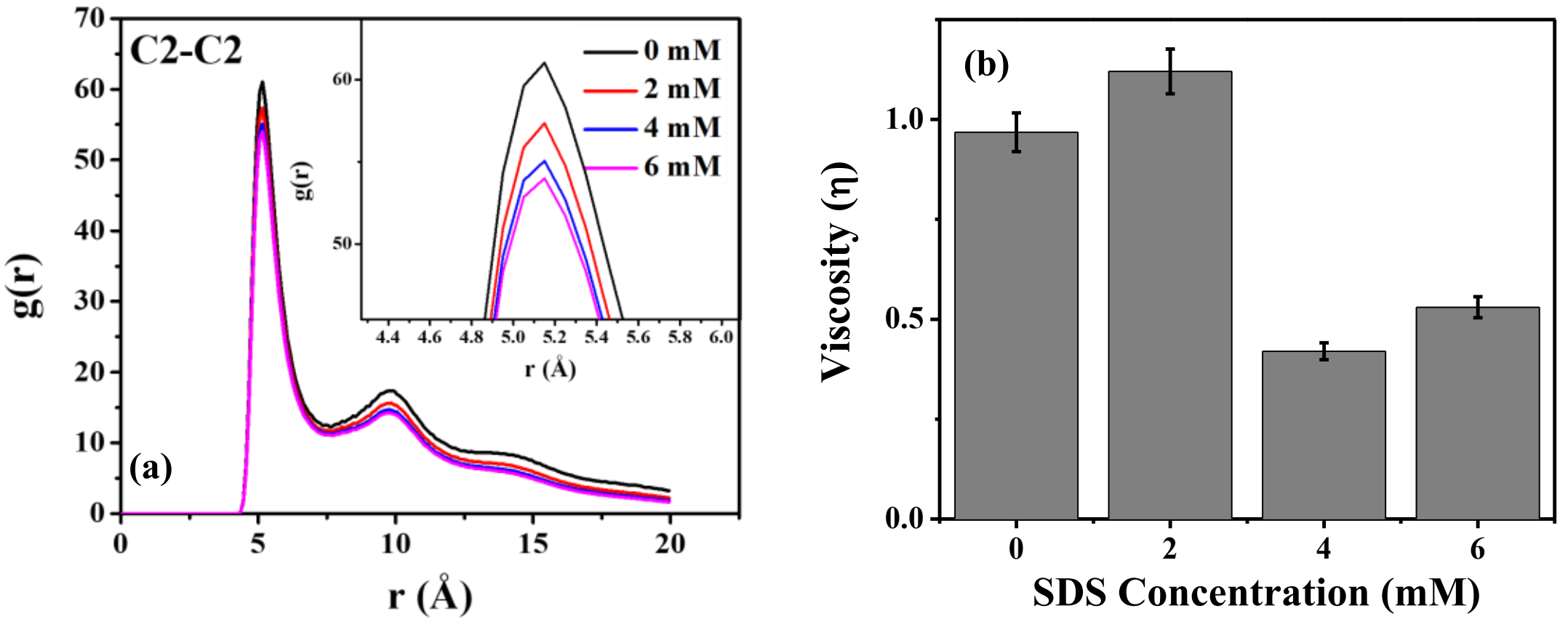

***Figure 7.*** ***(a)*** *Radial distribution function (RDF) between IL tail beads (C2-C2) at various SDS concentrations.* ***(b)*** *Zero-shear viscosity of the IL–SDS system as a function of SDS concentration.*

The incorporation of SDS into $[OMIM]^{+}[Cl]^{-}$ micelles leads to the formation of stable mixed micelles, primarily driven by electrostatic and hydrophobic interactions. Both experimental and simulation results indicate deviations from ideal micellization, as reflected in CMC values differing from Clint's predictions and negative interaction parameters (β) from the Rubingh model. The experimental results point to synergistic interactions facilitating the mixed micelle formation. A particularly intriguing feature is the non-monotonic behaviour observed at 4 mM SDS, where micellar stability decreases sharply. This instability is reflected across multiple observables: the interaction parameter β becomes less negative, viscosity reaches a minimum, SDS exhibits enhanced mobility within micelles, and IL–IL tail–tail correlations weaken significantly. At this intermediate concentration, SDS molecules may only partially insert into the micellar interface, leading to irregular charge distribution, which disrupt cohesive interactions within the aggregate. This convergence of indicators suggests a transitional micellar regime. Beyond this threshold, at higher SDS concentrations (e.g., 6 mM), the system appears to reorganize into more stable mixed micelles, where improved surfactant distribution and tail alignment restore structural integrity.

Micelle size distribution analysis (Figure S2) provides clear structural evidence of SDS modulating the micellar organization in the IL aggregates. At 0 mM SDS, the distribution is skewed toward smaller aggregates, with a sharp peak below 300 CG beads and a long tail extending to larger sizes. This indicates that $[OMIM]^{+}[Cl]^{-}$ forms numerous small micelles along with a few larger aggregates, suggesting a broad and polydisperse micellar population. Upon addition of SDS at 2 mM, the distribution remains skewed but becomes narrower and more centred around ~100-200 beads, indicating the formation of more compact and uniform mixed micelles. However, at 4 mM SDS, the distribution broadens again and shows increased variability in micelle sizes, reflecting micellar heterogeneity and transient destabilization, consistent with the observed minimum in viscosity and weaker IL–IL tail correlations. Finally, at 6 mM SDS, the distribution sharpens and centres between 150–250 beads, indicating reorganization into structurally stable and uniform micelles. The evolution of micelle size distribution with SDS concentration reinforces the non-monotonic behaviour observed, and highlights a critical concentration window around 4 mM where micellar architecture is disrupted before restabilizing at higher surfactant content. Although, the current study has extended the CG-MD to establish the ambiguity in micellar structure at 4 mM SDS concentration, but a detailed all atomistic molecular dynamics simulations is required.

## 5. Conclusion

The study explores the effect of an anionic surfactant, sodium dodecyl sulfate (SDS), on the micellization behaviour of the ionic liquid 1-methyl-3-octyl imidazolium chloride ($[OMIM]^{+}[Cl]^{-}$) in aqueous solution. Using a combination of surface tension measurements, existing theoretical models and coarse-grained molecular dynamics simulations, the formation of stable mixed micelles across a range of SDS concentrations has been depicted. Thermodynamic analysis using Clint and Rubingh models confirmed the non-ideal nature of IL–surfactant mixing, with strong synergistic interactions driving micelle formation. The results revealed a non-monotonic trend in micelle stability. These effects are further supported by consistent changes in viscosity, micellar composition, radius of gyration, and SDS mobility observed across experiments and simulations.

**Conflict of Interest**
There are no conflicts to declare

**Acknowledgement**
Devansh Kaushik, Syed M. Kamil, and Sajal K. Ghosh acknowledge the Shiv Nadar Foundation for supporting the research.

# Supplementary Information

## Concentration-Dependent Restructuring of Ionic Liquid Micelles Induced by SDS

Devansh Kaushik[a*], Sajal K. Ghosh[a], and Syed M. Kamil[a*]

[a]Department of Physics, School of Natural Sciences, Shiv Nadar Institution of Eminence, NH 91, Tehsil Dadri, Uttar Pradesh -201214, India

### *1. The system details for the simulation setup*

| IL Concentration | IL Molecules | SDS concentration | SDS Molecules | Embedded water | Water | |
|---|---|---|---|---|---|---|
| | | | | | P4 | BP4 |
| 350 mM | 1008 | 0 mM | 0 | 6048 | 34639 | 3849 |
| | | 2 mM | 6 | 6084 | 34631 | 3847 |
| | | 4 mM | 12 | 6120 | 34623 | 3847 |
| | | 6 mM | 18 | 6156 | 34616 | 3846 |

### *2. Analysis of critical micelle concentration using the surface tension measurement*

For each experiment, the surface tension of IL/ surfactant solution with highest concentration was measured initially. The solution was subsequently diluted, and the ST was measured at each dilution. The values of surface tension as a function of concentration have been plotted in Figure 5.2. From the figure the CMC estimation of pristine IL and surfactant aqueous solution remains straightforward. Similarly, for the mixed IL-SDS system, as one starts from the highly concentrated solution, which is progressively diluted. Under this protocol, the aggregates formed at high concentrations remain intact during dilution until their stability threshold is crossed. Consequently, when the dilution starts from the highest concentration, the initial decay in surface tension does not reflect the onset of micellization, but rather the progressive destabilization of pre-formed micelles. The plateau that follows corresponds to a coexistence regime where micelles persist in a reorganized form and, are rich in SDS as the system does qualify the thresholds for IL rich micelles, which pushes the air-water interface to be rich in ionic liquids. The sharp increase in surface tension observed on the right-hand side therefore marks the concentration below which IL rich micelles can no longer survive. Since the critical micelle concentration (CMC) is, by definition, the threshold separating monomeric surfactant from micellar aggregates, the right-side rise is the most appropriate indicator of the mixed CMC to probe the effect of SDS on IL rich micelles, in this dilution-based experimental design. Further moving to higher dilution regime, the second rise reflects the formation of SDS rich micelles, where IL reduces the CMC of pristine SDS system.

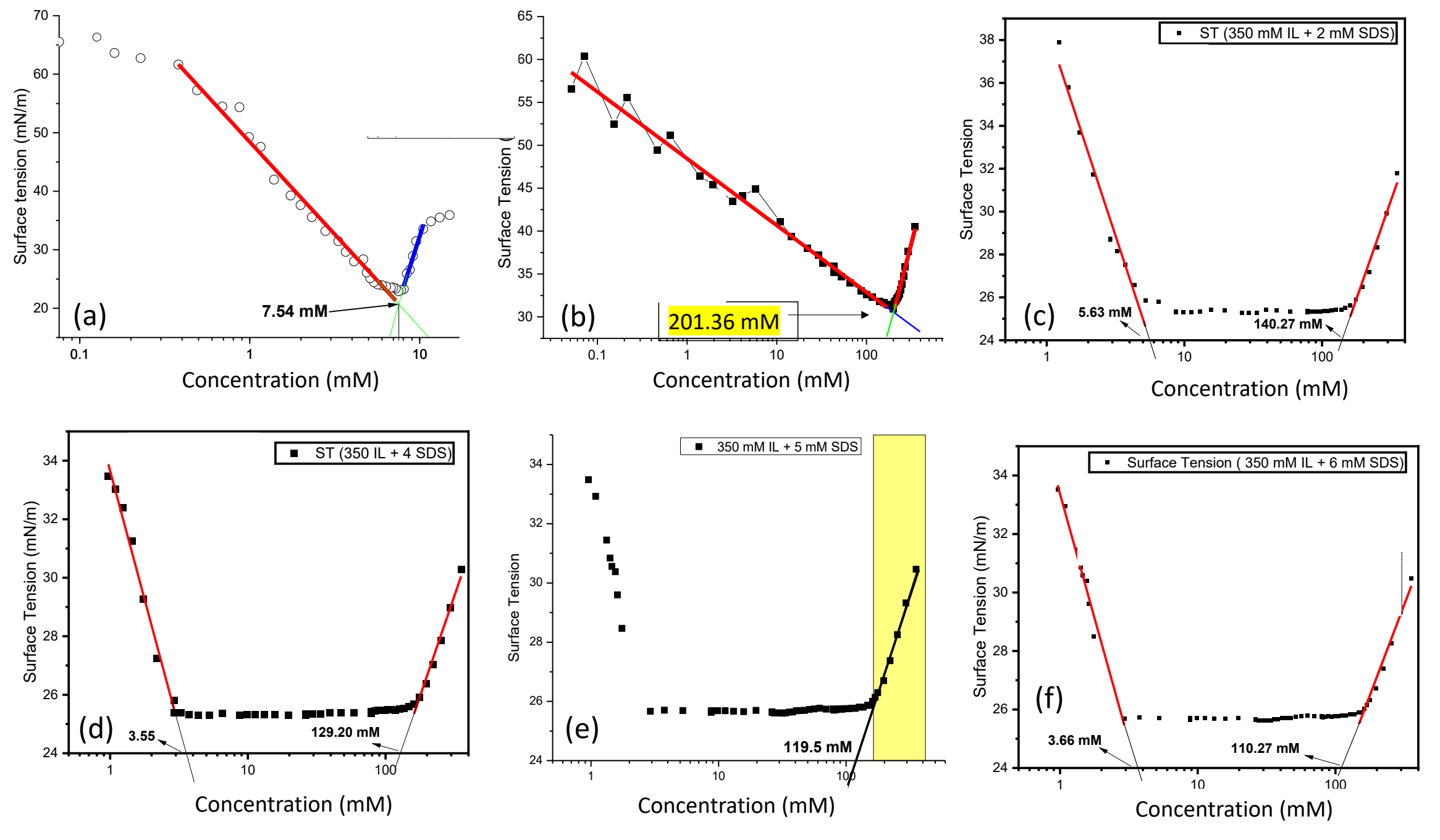

***Figure S1.*** *The calculation of CMC values using the surface tension methods for (a) SDS, (b) $[OMIM]^+[Cl]^-$, (c-e) 2, 4 and, 6 mM SDS in 350 mM IL.*

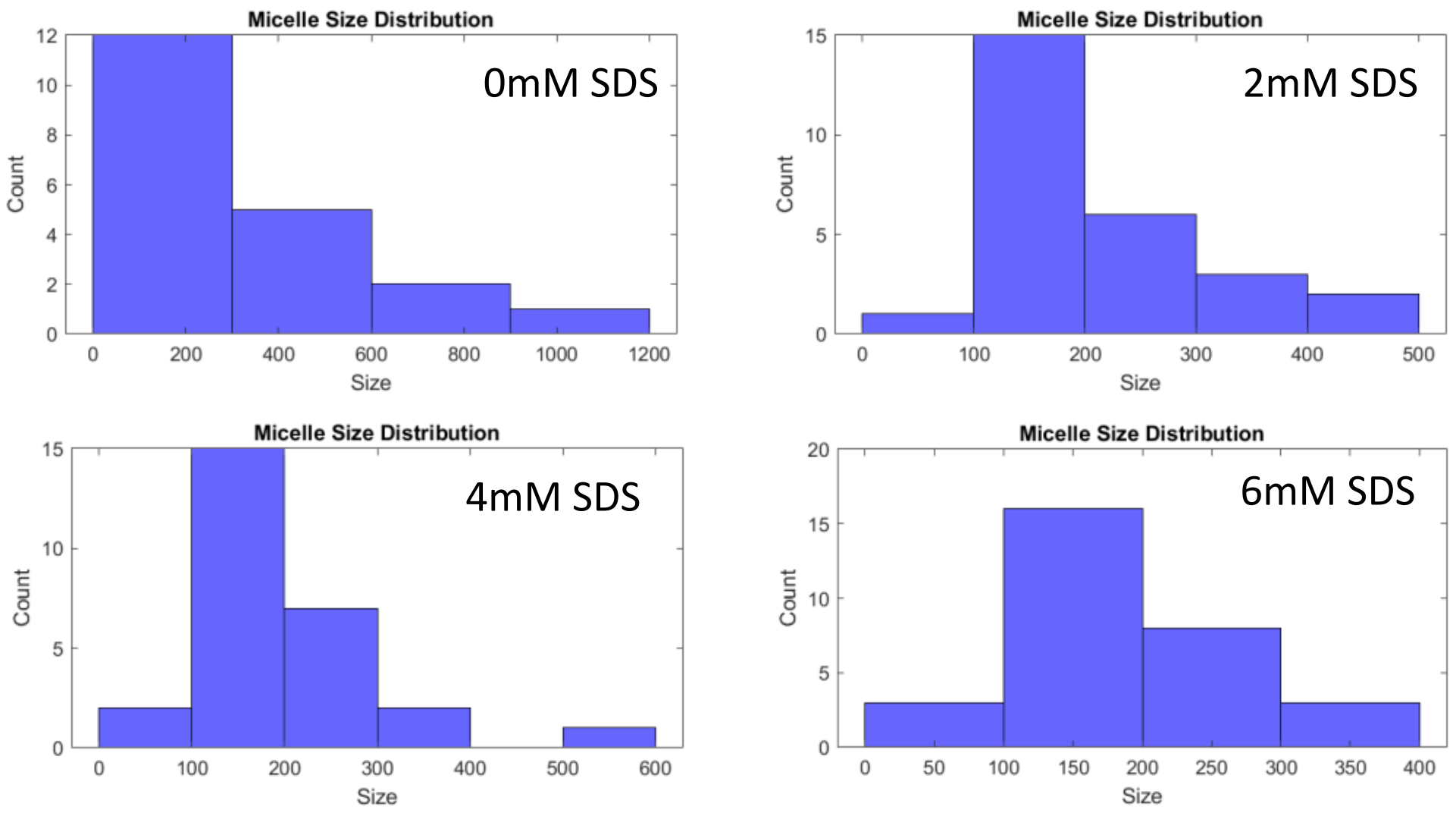

***Figure S2.*** *The cluster size distribution analysis for different SDS concentration in the IL rich micelles.*

### 3. Calculation for diffusion ($l_d$) of the SDS molecule in the IL micellar structure.

The local motion of the SDS molecule within the IL-rich micellar aggregates was analyzed from coarse-grained trajectories using Ovito (Open Visualization Tool). For each concentration, the trajectory of the terminal tail bead (T3) of a representative SDS molecule was tracked for 200 ps after equilibration. The center of mass (COM) of the trajectory was calculated as

$$r_{COM} = \frac{1}{N}\sum_{i=1}^{N} r_i$$

Where, $r_i$ is the instantaneous position of the bead in frame i, and N is the total number of frames analysed. The spread of the trajectory about its COM was then quantified through the radius of gyration, defined as

$$l_d = R_g = \sqrt{\frac{1}{N}\sum |r - r_{COM}|^2}$$

in the present work, this $R_g$ was interpreted as the diffusion length ($l_d$) of the SDS molecule, since it represents the characteristic spatial extent explored by the molecule over the observation window. Outlier displacements (>75 Å) were omitted to exclude rare inter-aggregate transitions. The obtained $l_d$ values, reflect the relative confinement of SDS motion inside the micellar domain.

### *4.* Uncertainty estimation for derived thermodynamic parameters.

The interaction parameter (β), micellar mole fraction ($X_1$), and activity coefficients ($\gamma_1$ for $[OMIM]^+[Cl]^-$ and $\gamma_2$ for SDS) were not directly measured but were obtained from experimentally determined critical micelle concentrations (CMCs) using Rubingh's regular solution theory. To quantify the uncertainty in these derived quantities, Monte Carlo error propagation was employed. The experimentallyf measured CMC values (CMC_mix, $CMC_1$, and $CMC_2$) were treated as normally distributed variables with standard deviations obtained from repeated surface-tension measurements. For each SDS concentration, a large ensemble ($10^4$ realizations) of CMC values was generated, and the Rubingh equations were solved numerically for each realization to obtain distributions of $X_1$, β, and γ. The reported values correspond to the mean of these distributions, while the error bars shown in the figures represent one standard deviation. This approach ensures that the uncertainties in β and activity coefficients faithfully reflect experimental variability in CMC determination and preserves the correlations inherent to the regular solution model.